\documentclass[%
aip, apl,
amsmath,amssymb,
reprint,%
]{revtex4-1}

\usepackage{graphicx}
\usepackage{dcolumn}
\usepackage{bm}
\usepackage{xcolor} 
\usepackage[utf8]{inputenc}
\usepackage[T1]{fontenc}
\usepackage{mathptmx}
\usepackage{etoolbox}

\makeatletter
\def\@email#1#2{%
 \endgroup
 \patchcmd{\titleblock@produce}
  {\frontmatter@RRAPformat}
  {\frontmatter@RRAPformat{\produce@RRAP{*#1\href{mailto:#2}{#2}}}\frontmatter@RRAPformat}
  {}{}
}%
\makeatother
\begin{document}

\preprint{AIP/123-QED}

\title{Spin Accumulation based deep MOKE Microscopy}

\author{J. C. Rodriguez E.}

\affiliation{%
Departamento Magnetismo y Materiales Magnéticos, Centro Atómico Bariloche, Gerencia de Física, Centro Atómico Bariloche, Av. Bustillo 9500, (8400), S. C. de Bariloche (RN), Argentina
}%
\affiliation{%
Instituto de Nanociencia y Nanotecnología (CNEA-CONICET), Nodo Bariloche, Av. Bustillo 9500, (8400), S. C. de Bariloche (RN), Argentina
}%
\affiliation{%
Instituto Balseiro, CNEA-UNCuyo, Av. Bustillo 9500, (8400), S. C. de Bariloche (RN), Argentina
}%

\author{H. Grisk}%

\author{A. Anadón}
\author{H. Singh}
\author{G. Malinowski}
\author{M. Hehn}
\affiliation{%
Université de Lorraine, CNRS, IJL, Nancy, F-54000, France
}%
\author{J. Curiale}
\affiliation{%
Departamento Magnetismo y Materiales Magnéticos, Gerencia de Física, Centro Atómico Bariloche, Av. Bustillo 9500, (8400), S. C. de Bariloche (RN), Argentina
}%
\affiliation{%
Instituto de Nanociencia y Nanotecnología (CNEA-CONICET), Nodo Bariloche, Av. Bustillo 9500, (8400), S. C. de Bariloche (RN), Argentina
}%
\affiliation{%
Instituto Balseiro, CNEA-UNCuyo, Av. Bustillo 9500, (8400), S. C. de Bariloche (RN), Argentina
}%

\author{J. Gorchon}
 \email{jon.gorchon@univ-lorraine.fr}
\affiliation{%
Université de Lorraine, CNRS, IJL, Nancy, F-54000, France
}%

\date{\today}
\begin{abstract}
Magnetic imaging techniques are widespread critical tools used in fields such as magnetism, spintronics or even superconductivity. Among them, one of the most versatile methods is the magneto-optical Kerr effect. However, as soon as light is blocked from interacting with the magnetic layer, such as in deeply buried layers, optical techniques become ineffective. In this work, we present a spin-accumulation based magneto-optical Kerr effect (SA-MOKE) microscopy technique that enables imaging of a magnetic thin-films covered by thick and opaque metallic layers. The technique is based on the generation and detection of transient spin-accumulations that propagate through the thick metallic layer. These spin-accumulation signals are directly triggered and detected optically on the same side, lifting any substrate transparency requirements. The spin-accumulation signals detected on a Cu layer decay with a characteristic length of 60 nm, much longer than the 12 nm optical penetration depth, allowing for detection of magnetic contrast with Cu capping layers up to hundreds of nm. This method should enable magnetic imaging in a wide-range of experiments where the surface of interest is covered by electrodes.
\end{abstract}

\maketitle
Magneto-optics is routinely used by many to characterize magnetic properties such as coercivity, anisotropy, or even the Dyalozhinkii-Moriya interaction \cite{Balk2017, Soucaille2016}. In particular, the spatial information conveyed by Kerr or Faraday microscopy is often critical to understand numerous experiments. Notable examples include domain wall motion in racetracks\cite{Emori2013,Yang2015,Ryu2013}, voltage controlled magnetic anisotropy \cite{ContePRM2018,Yang2023,Fillion2022}, skyrmion nucleation and propagation \cite{Jiang2015, Fillion2022,Yang2024} and all-optical magnetization switching \cite{Stanciu2007,Igarashi2023,Peng2023}. Moreover, the use of ultrashort laser pulses in a pump-probe setting extends the time-resolution down to the femtosecond\cite{Beaurepaire1996}, all while conserving the same spatial information \cite{Hashimoto2014,Bigot2013,Remy2023}. However, if an opaque layer or electrode covers the magnetic material \cite{Chiba2011,Geiskopf2025}, magneto-optics become ineffective. Nevertheless, pump-probes studies of magnetic layers in contact with thick and opaque metallic layers have revealed the possibility of detecting spin accumulations on the surface of the non-magnetic layer \cite{Choi2014,Choi2015,Anadon2025}, which are due to the laser-induced ultrafast demagnetization of the ferromagnet. Therefore, in such experiments, the magnetic information from the distant magnet, can be partially conveyed to the surface of the normal metal beyond the optical penetration depth. Such remote sensing is reminiscent of the field of picosecond acoustics, where time-domain thermoreflectance experiments are used to probe deeply buried layers by measuring the transmission and reflection of laser-triggered acoustic pulses \cite{Thomsen1984,Gusev2023}.

In this letter, we demonstrate optical imaging of magnetic domains covered by Cu layers up to 140 nm thick. Our technique, spin accumulation magneto-optical Kerr effect (SA-MOKE) imaging, combines a pump-probe magneto-optical setup with a scanning stage to measure the transient spin-accumulation induced by the pump at the surface of the Cu layer, as a function of position. The SA-MOKE technique results in high contrast images with a domain structure unperturbed by the laser irradiation.

We grew a batch of three samples by DC magnetron sputtering onto transparent sapphire substrates. The multilayer structures are the following: substrate//Ta(3)/Cu(5)/ [Ni(0.7)/Co(0.2)]{$_{ \times 4}$}Cu($t_\text{Cu}$)/Al(3) (thickness in nm). The top Cu layers have a variable thickness $t_\text{Cu}$ ranging from 10 to 30 nm, 20 to 60 nm and 50 to 150 nm \cite{Anadon2025}. The top Al(3) cap layer is expected to be naturally passivated.
An additional wedge sample, with $t_\text{Cu}$ ranging from 50 to 150 nm and [Ni(0.45)/Co(0.2)]{$_{ \times 4}$} as the ferromagnetic multilayer, was specially  prepared for domain imaging. 
All samples present dominant perpendicular magnetic anisotropy. For clarity, we will use the following notation to refer to the ferromagnetic Ta/Cu/[Co/Ni] part of the stack as the FM layer, and to the top Cu/Al as the Cu layer.

Time-resolved magneto-optical Kerr effect (TR-MOKE) experiments are performed using the setup presented in Ref. \cite{Anadon2025}. The setup uses a Ti:Sa femtosecond laser system with a repetition rate of 80 MHz and a center wavelength of 800 nm. For these experiments, the probe beam is frequency doubled to 400 nm, due to its higher sensitivity to spin accumulation\cite{Singh2025}. The beams were focused to spot sizes (FWHM) of approximately 50 $\mu$m for the pump and 5 $\mu$m for the probe. The XY position of the sample was then controlled via two linear stepper motor stages (ZST225B and ZFS25B from Thorlabs).

The microscopes used to acquire magnetic domain images through a sapphire substrate were a commercial wide field MOKE (WF-MOKE) microscope from Evico Magnetics\cite{EvicoMagnetics} and also a home-made polar MOKE microscope\cite{Jhuria2020}. The WF-MOKE images will be compared with the imaging results obtained by SA-MOKE technique.

\begin{figure}
\centering
                \includegraphics[width=0.47\textwidth]{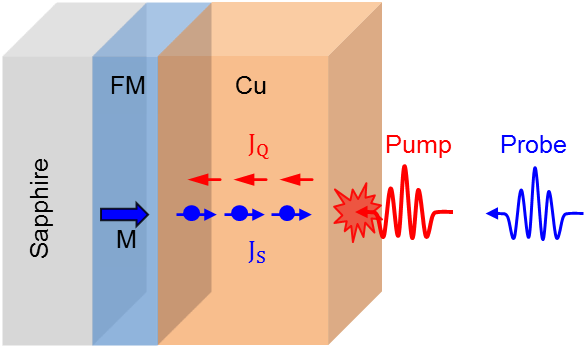} 
                \caption{\label{fig:Scheme}
                \textbf{Conceptual schematic of SA-MOKE}. The pump laser pulse (red wavy arrow) is incident on the Cu side and generates an electronic heat current ($\mathrm{J_{Q}}$) which flows up to the ferromagnetic (FM) layer and induces its demagnetization. The demagnetization in turn generates a spin current ($\mathrm{J_{S}}$) back into the Cu layer that flows to its surface. Finally, the probe laser pulse (blue wavy arrow) measures the magneto-optical Kerr rotation induced by the spin accumulation at the surface of the Cu layer. This signal is proportional to the magnetization of the deeply buried FM layer.
                }
\end{figure}

In Fig.~\ref{fig:Scheme} we show the conceptual principle of SA-MOKE. A pump laser pulse (in red) excites the Cu-side and generates hot electrons that travel towards the FM. This electronic heat current $J_Q$ leads to ultrafast demagnetization of the FM layer \cite{Bergeard2016}. The demagnetization, in turn, injects a spin current $J_S$ back into the Cu layer \cite{Choi2014}. When the Cu layer is sufficiently thick ($>~60$ nm), the probe (in blue), incident on the Cu-side, detects a magneto-optical signal which is attributed solely to spin accumulation \cite{Anadon2025, Choi2014}. The amplitude and sign of the detected signal are therefore directly proportional to the demagnetization of the layer, which itself is proportional to the magnetization $\vec{M}$. 
Finally, in order to create a full image of the magnetic topography of the buried FM layer, we just need to scan the surface with the XY stages.

\begin{figure}[ht!]  
\centering
                \includegraphics[width=0.48\textwidth]{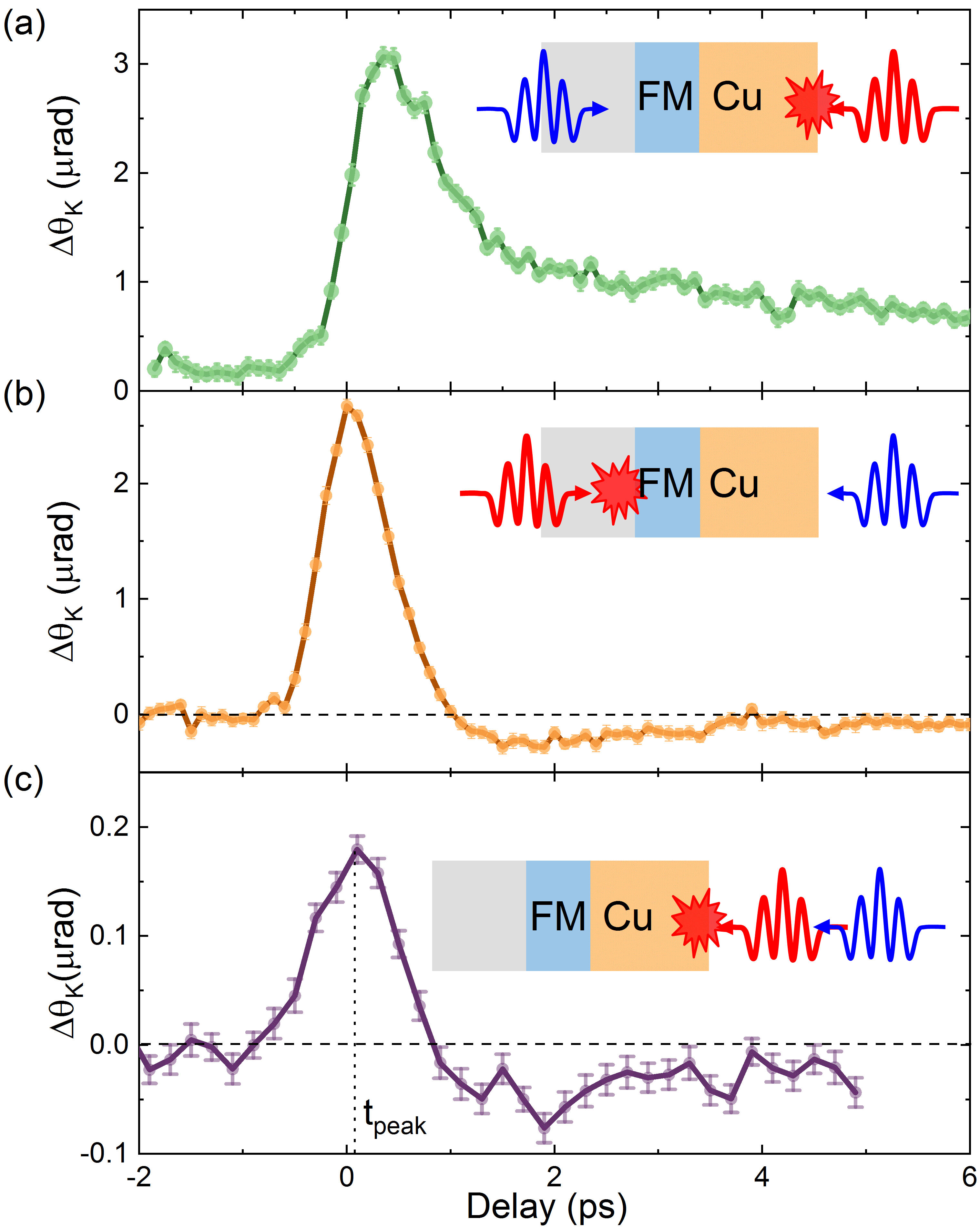}      
                \caption{\label{fig:Setup}
                \textbf{Time-resolved magneto-optical Kerr responses in the [Co/Ni]/Cu structure in various configurations. 
                } \textbf{a)} Demagnetization of the FM layer due to hot electrons from the Cu layer. \textbf{b)} Spin accumulation in the Cu layer due to the optical heating of the FM layer. \textbf{c)} Spin accumulation in the Cu layer due to electronic heating of the FM layer. Schematics on the upper-right show the experimental configurations of pump (red pulse) and probe (blue pulse) in each case. We used incident pump fluences of 0.2 and 0.1 mJ/cm$^2$ for the experimental configurations \textbf{a-b} and \textbf{c}, respectively. The measurements were performed on a Cu-100 nm layer. The time zero is not comparable because the measurements were performed in different experimental conditions. Integration times and/or number of averaged scans are different in panels (a), (b) and (c). The standard error for measurements at each time delay is represented by the size of the data point markers (a-b) or error bars (c).}
\end{figure}

We first perform TR-MOKE experiments to check the dynamics of the magnetization in the FM layer and the spin accumulation in the Cu layer.
In order to avoid non-magnetic contributions, we measure and subtract the pump-induced Kerr rotation in both out-of-plane magnetization configurations\cite{Anadon2025}. In Fig.~\ref{fig:Setup}a we plot the typical transient Kerr rotation signal, as detected on the FM-side, induced by the excitation of the Cu layer. We observe a classical ultrafast demagnetization curve (sign of $\Delta\theta_k$ depends on the magneto-optical indices) due to the electronic heat current arriving from the Cu layer \cite{Bergeard2016}. In Fig.~\ref{fig:Setup}b we plot the signals obtained in the opposite configuration: pumping the FM and probing on Cu-side. In this situation, the FM is excited and demagnetized directly by the pump beam, which leads to spin injection into the Cu layer, and to a detectable spin accumulation \cite{Anadon2025,Choi2014}. Note that the shape of the spin accumulation is now slightly bipolar, as it is related to the time-derivative of the demagnetization\cite{Anadon2025,Choi2014} (visible in Fig.~\ref{fig:Setup}a). Here, we focus on the Kerr rotation, as the Kerr ellipticity in Cu due to spin-accumulation is quite smaller at a 400 nm wavelength \cite{Singh2025}. Finally, shown in Fig.~\ref{fig:Setup}c, we combine the latter two experiments and detect spin accumulations on the Cu surface which are generated via hot electron currents, as also schematized in Fig.~\ref{fig:Scheme}. The magneto-optical signal has a similar behavior with the previous experimental condition; however, the signal decreases by roughly an order of magnitude. This is not surprising as the thick Cu absorbs very little energy compared to the FM, and not all of it reaches the FM. Interestingly, the amplitude and sign of the peak in Fig.~\ref{fig:Setup}c will be directly proportional to the local magnetization, below the Cu. In the following, we will monitor this peak-value, by fixing the time delay at $\text{t}_{\text{peak}}$, to perform SA-MOKE imaging.

 \begin{figure*}[ht!]
\centering
                \includegraphics[width=0.8\textwidth]{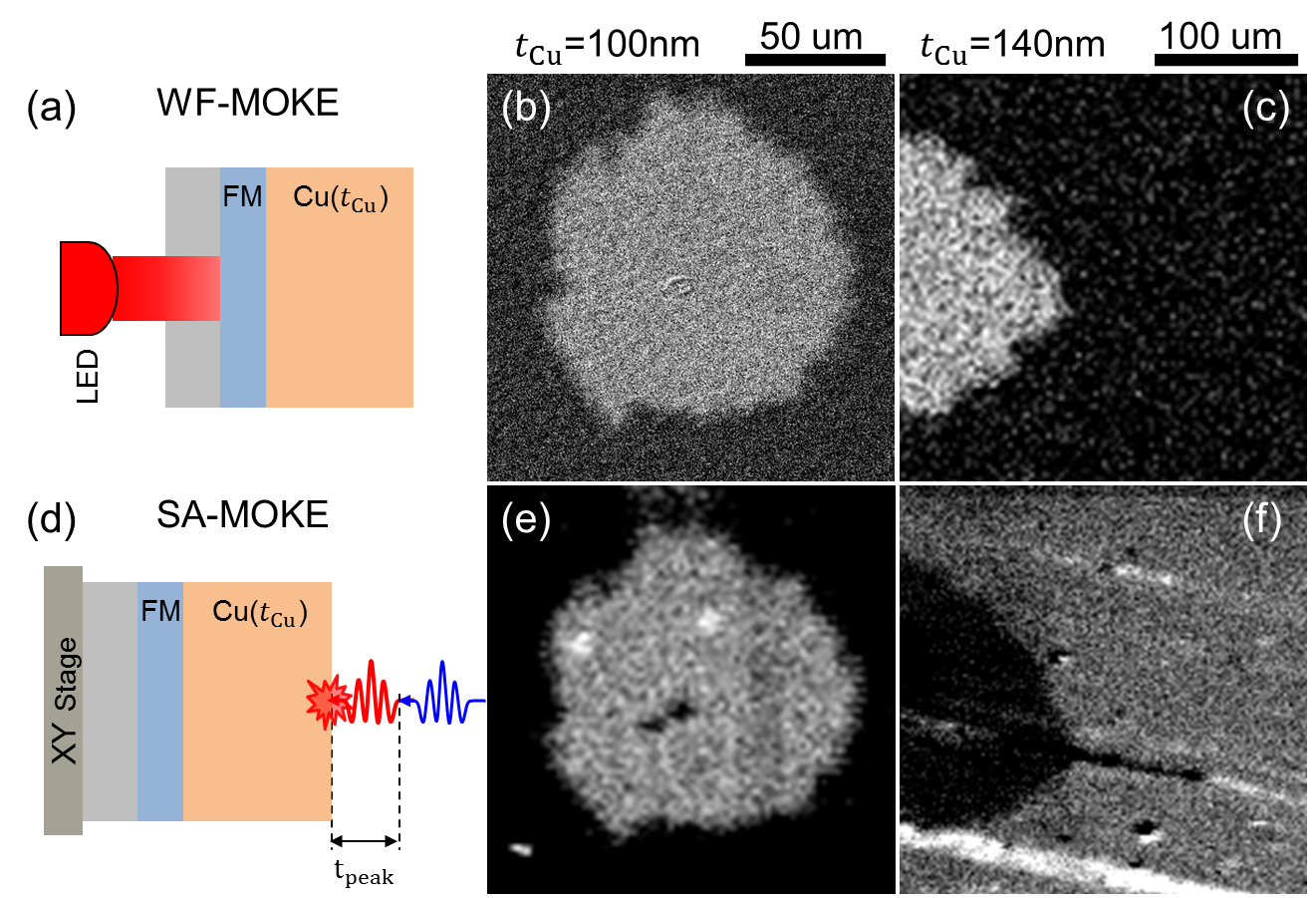}

                \caption{\label{fig:Imaging}
                \textbf{Spin accumulation based MOKE imaging of magnetic domains through a thick Cu layer.} \textbf{a)} Schematic representation of conventional polar wide-field (WF) MOKE microscopy, measuring the magnetic layer directly through the substrate. Magnetic domain images obtained for Cu thickness of \textbf{b)} 100 nm  and \textbf{c)} 140 nm by WF-MOKE microscopy. \textbf{d)} Schematic representation of the SA-MOKE technique to image magnetic domains. Magnetic domain images obtained for Cu thickness of \textbf{e)} 100 nm and \textbf{f)} 140 nm by SA-MOKE. We used pump fluences of 0.1 and 0.2 mJ/cm$^2$ for the scanning TR-MOKE \textbf{e} and \textbf{f}, respectively. All measurements were carried out at room temperature. 
                }
\end{figure*}

We nucleate magnetic domains in the Co/Ni film using an external magnetic field, and monitor them through the transparent substrate with a conventional WF-MOKE microscope\cite{EvicoMagnetics, Jhuria2020}, as depicted in Fig.~\ref{fig:Imaging}a. Examples are shown in Fig.~\ref{fig:Imaging}b-c. where the black/gray contrast represents domains with opposite out-of-plane magnetization orientations. We then measure the peak spin accumulation signal, at a fixed time-delay $\text{t}_{\text{peak}}$, and scan the XY position, as depicted in Fig.~\ref{fig:Imaging}d. The resulting images obtained when probing on 100 nm and 140 nm of Cu are shown in Fig.~\ref{fig:Imaging}e and f, respectively. 
The SA-MOKE images exhibit excellent quality and contrast, comparable to the WF-MOKE results, despite the presence of 100 – 140 nm of Cu blocking the incident light. Moreover, the perfect match between the magnetic domain patterns obtained by both methods indicates that the excitation pump pulses do not alter the domain structure. This behavior is a direct consequence of the small absorbed energy of only a few $\mu$J/cm$^2$ (with the optical absorption estimated to be about 3$\%$ using a generalized transfer matrix model), which induces only a minor demagnetization.
The spatial resolution of the SA-MOKE images is determined by a number of factors. The probe spot size and the step size of the motorized stage determine the minimum pixel size. In our case, we reached the diffraction limit of the blue illumination, achieving a resolution comparable to that of the WF-MOKE. Generally, the pump can excite a large area, without compromising the resolution, as electron and spin transport can be considered 1D (out-of-plane) in such thin film systems. However, for very thick layers (comparable to the probe spot size), lateral spin transport could start to play a role reducing the overall resolution.

Although the contrast and spatial resolution of the SA-MOKE images are excellent, it is worth noting that the experiments took much longer than their WF-MOKE counterparts — for example, 17 hours in Fig.~\ref{fig:Imaging}e. Remarkably, in some cases — for example, in Fig.~\ref{fig:Imaging}e — we observed additional brighter and/or darker spots in the SA-MOKE microscopy images. These small fluctuations are not related to the magnetization, as they were absent when the domain structure was re-examined using the WF-MOKE microscope. These features are likely due to surface imperfections or residual particles on the Cu layer. After careful cleaning with acetone and ethanol, different pattern patterns were observed or no pattern at all. Although detailed investigation would be required to fully elucidate the origin and relevance of these patterns, such a study falls outside the scope of this work.

In the following paragraphs, we analyze the signal-to-noise ratio and discuss its impact on the required experimental integration time for different Cu thicknesses.

\begin{figure}[htb]
\centering 
                \includegraphics[width=0.47\textwidth]{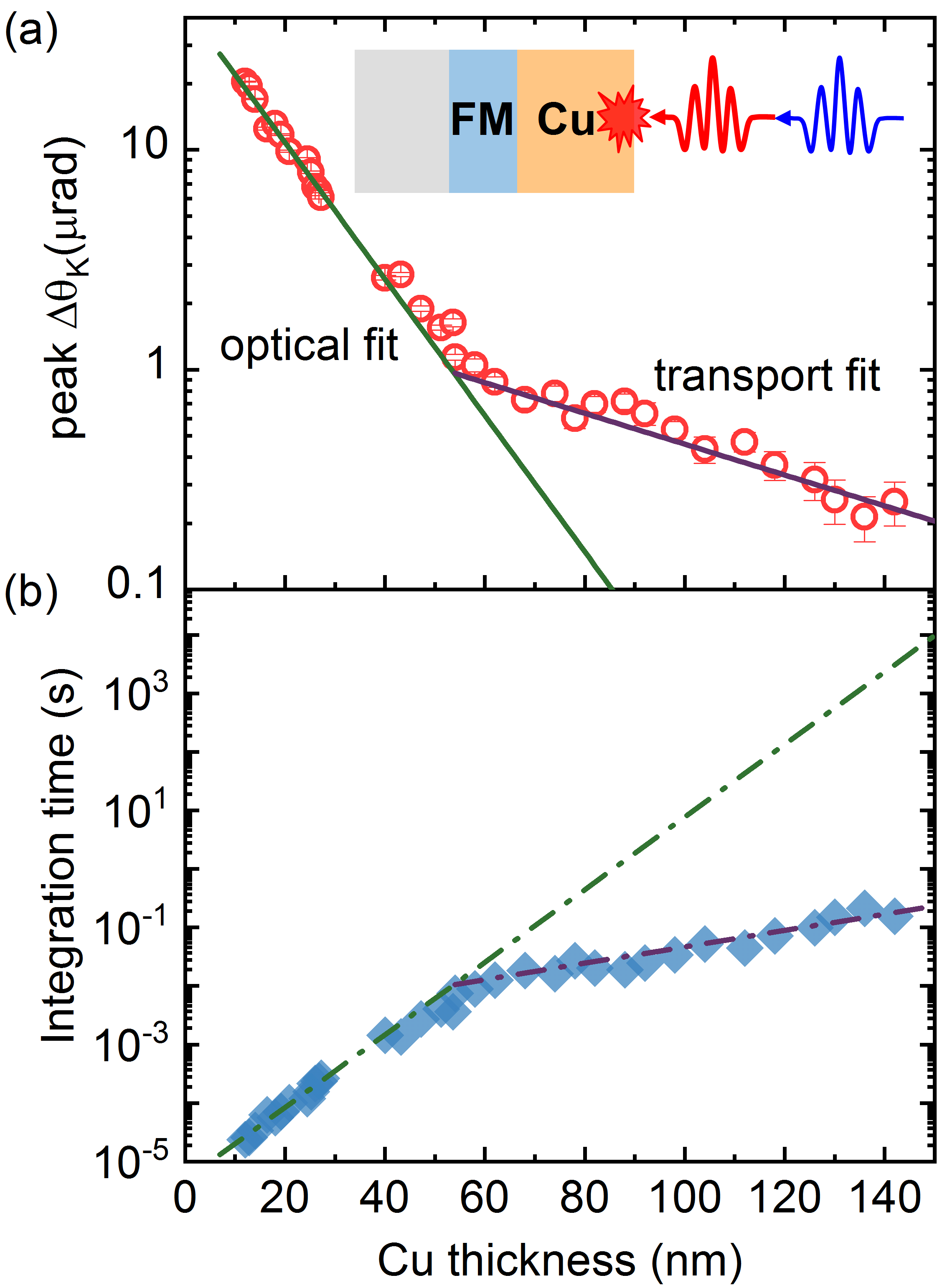}
                \caption{\label{fig:KerrSignal}
                \textbf{Cu thickness dependence on Kerr rotation and integration time for signal-to-noise ratio of 1.} \textbf{a)} The peak of Kerr rotation as a function of Cu thickness for CoNi/Cu structure at fluence of 0.2 mJ/cm$^2$. The peaks of Kerr rotation, associated with demagnetization and spin current, are plotted using red open circles. The red and purple lines are exponential fits for lower and higher Cu thickness. The error bars are the standard error. \textbf{b)} Integration time to reach a signal-to-noise of 1, calculated from the peaks of Kerr rotation as a function of Cu thickness, is plotted using blue solid diamonds. The green and purple dash dotted lines correspond to integration time estimated from the optical and transport exponential fits, respectively.}
\end{figure}

Figure~\ref{fig:KerrSignal}a shows the peak of the Kerr rotation signal for Cu thicknesses ranging from 15 nm to 140 nm. Two distinct linear regimes can be identified in the logarithmic plot, indicating the presence of two exponential decay functions.
As we previously discussed, for thin Cu layers ($t_\text{Cu} < 20$ nm), light can penetrate the Cu layer, and the detected signal is therefore associated with the magnetization of the Co/Ni layer (i.e., the ultrafast demagnetization), as shown in Fig.~\ref{fig:Setup}a. In contrast, for thick Cu layers ($t_\text{Cu} > 100$ nm), as discussed earlier in Fig.~\ref{fig:Setup}c, the peak clearly originates from spin accumulation.

Taking this into account, the observed behavior for $t_\text{Cu} \lesssim 50$ nm (green fit) can be mostly attributed to the optical penetration depth of the light, and thus to the reduced direct magneto-optical sensitivity to the FM layer. Additionally, below $\sim$ 50 nm the thickness dependence of the absorption of light and heat transport in the Cu will also play a role. From the fit we extract a characteristic length of 14 nm, a value close to double the optical penetration depth in Cu, which considers the back-and-forth trip of the electro-magnetic wave \cite{Anadon2025, Hubert1998}. The second regime (fitted in purple), corresponds to the reduction of the spin accumulation signal. This reduction is primarily due to two effects: electron-phonon scattering in the Cu which reduces the amount of heat that reaches the FM, and spin-flip relaxation in the Cu which limits the spin-diffusion length. For this reason, the characteristic length of 62 nm extracted from the fit is shorter than the typical spin-diffusion length of 120 nm measured in the same structures\cite{Anadon2025}.

 The signal-to-noise ratio (SNR) of our experiment is defined as the ratio between the peak $\Delta \theta_K$ and the root-mean-square noise angle, $\Delta \theta_\text{Noise}^{\text{rms}}$. The noise amplitude spectral density of the experiment (NASD) was estimated to $0.2~\mu rad/\sqrt{Hz}$ by measuring the noise in a single scan and estimating the bandwidth from lock-in filters and integration times. The noise in a specific measurement is then given by $\Delta \theta_{Noise}^{rms}=NASD/\sqrt{2T}$. We can therefore estimate the required integration time of a measurement to reach a SNR of 1, for all Cu thicknesses, as shown in Fig.~\ref{fig:KerrSignal}b. Extrapolating the direct magneto-optical data (green fit), we estimate an integration time of about 2000 s for a Cu thickness of 140 nm. In contrast, the spin-accumulation signal can be measured in only ~0.2 s, representing an improvement by a factor of roughly 10,000. For $t_\text{Cu} = 140$ nm, acquiring a 100×100 pixel image with a SNR = 1 takes about 30 minutes with SA-MOKE, compared to roughly 230 days with direct magneto-optical methods.

The presented method offers a unique advantage to probe magnetization when metal layers > 50 nm need to be deposited on top of the magnetic structure. However, SA-MOKE still remains a very \textit{slow} technique at the moment due to the limited signal-to-noise ratio and scanning method. In fact, due to its scanning nature, a tradeoff trilemma exists between the field-of-view, the resolution, and the experiment duration. In order to improve on these limitations, various options can be envisioned: First, a stronger excitation of the magnet would result in large spin accumulations, and therefore faster measurements. This can be attained by using higher-power laser systems, tuning the wavelength for better absorption or even depositing some dielectric layers on top of the Cu to increase the energy absorption via cavity or anti-reflection effects. Nevertheless, the energy deposited by the laser should remain low enough so that the magnetic patterns are not affected by the heat. Second, the probe interaction can be optimized. Here again, a selection of the proper wavelength and/or extrinsic resonance effects can be explored. Additionally, the light-metal (here Cu) could be capped by a material which enhances the magneto-optical sensitivity (ex: Au \cite{Choi2018}). Finally, other light metals other than Cu could be envisioned, to increase either the attenuation lengths or the magneto-optical signals. According to the literature, Ag or Au could also be good candidates \cite{Choi2014PRB,Anadon2025} which would allow this technique to work with the most common types of electrodes. 

We note that SA-MOKE does not require specific substrates to work, but substrates with good thermal conductivity (saphire, MgO, semiconducting or metallic substrates...) are preferable to keep the static heating low. In practice, static heating is rarely a problem, as the thick metallic layer acts as a naturally efficient heat sink. We have demonstrated this technique for imaging, but magnetization hysteresis curves are equally readily accessible. Additionally, this technique can be extended to longitudinal MOKE probing configurations to potentially monitor all 3 components of the magnetization. Although a femtosecond laser has been used in this work, the technique should also work with commercially available few-picosecond-wide lasers. As a final note, the method could potentially also be adapted into a full-field, laser illuminated pump-probe microscopy setup \cite{Remy2023} exploiting the same spin-accumulation signals. Such evolution could potentially speed up the imaging drastically.

To conclude, we optically induce and detect spin accumulations on the surface of a Cu layer, which are directly linked to the magnetization profile of a ferromagnet sitting below. We show that these signals can then leveraged to image magnetic domain structures remotely even through thick non-magnetic layers up to 140 nm in thickness. The obtained SA-MOKE images are of comparable quality to those obtained with a classical wide-field (WF) MOKE microscope. We also systematically measure the peak of Kerr rotation as function of Cu thickness to reveal two distinct exponential decays: the first linked to the optical penetration depth in Cu, and a second one related to spin and heat transport. SA-MOKE should enable verification of magnetic properties and domain configurations in certain situations where no conventional probes can reach the target magnet, such as when using opaque electrodes to study voltage-controlled magnetic anisotropy effects or MTJ sensors.


\begin{acknowledgments}
This work was supported by the France 2030 government investment plan managed by the French National Research Agency under grant references PEPR SPIN – TOAST ANR-22-EXSP-0003 and SPINMAT ANR-22-EXSP-0007. This work was also partly supported by the french France 2030 program «Initiative d'Excellence Lorraine (LUE)», reference ANR-15-IDEX-04-LUE. We also thank EU-H2020-RISE project Ultra Thin Magneto Thermal Sensoring ULTIMATE-I (Grant ID. 101007825).
\end{acknowledgments}

\section*{Declarations}
\begin{itemize}
\item Competing interests: The authors declare no competing interests.
\item Availability of data: Data are available upon reasonable request to the corresponding author.
\end{itemize}

\section*{Author Contributions}
J.G. designed the experiments. J.G. and J.C. supervised the study. A.A. and H.S. built and optimized the optical setup and obtained preliminary data. J.C.R.E. and H.G. performed the magneto-optical experiments. M.H. performed the wedged sample deposition and optimized the magnetic properties. J.C.R.E., H.G., J.C. and J.G. analyzed the data with support from M.H. and G.M. J.C.R.E., J.C. and J.G. wrote the manuscript with input from all the authors.

\nocite{*}
\section*{References}\label{sec12}
\providecommand{\noopsort}[1]{}\providecommand{\singleletter}[1]{#1}%

\end{document}